\documentclass[aps,prl,reprint,superscriptaddress,eprint,nourl,nolongbibliography]{revtex4-2}

\usepackage[centertags]{amsmath}
\usepackage{amsfonts}
\usepackage{amssymb}
\usepackage{amsthm}
\usepackage{amscd}              

\usepackage{natbib}
\usepackage{bookmark}
\usepackage{url}
\usepackage{textcase}
\usepackage{bm}
\usepackage[pdftex]{graphicx} 
\graphicspath{{figures/}}

\usepackage{array}
\usepackage{color}
\usepackage{xcolor}
\usepackage{colortbl}

\usepackage[utf8]{inputenc}
\usepackage[T1]{fontenc}
\usepackage{lmodern}
\usepackage{bbold}
\usepackage{float}
\usepackage{microtype}
\usepackage{marvosym}

\newcommand{\ee}{\mathrm{e}}
\newcommand{\ii}{\mathrm{i}}

\newcommand{\ER}{E_\mathrm{R}}%
\newcommand{\wL}{\omega_\mathrm{L}}%
\newcommand{\kL}{k_\mathrm{L}}%
\newcommand{\bmkL}{\bm{k}_\mathrm{L}}%
\newcommand{\omegaL}{\omega_\mathrm{L}}%
\newcommand{\deltaL}{\delta_\mathrm{L}}%
\newcommand{\tnn}{t_{\mathrm{nn}}}

\begin{document}

\title{Many-Body Dephasing by Hole Motion in a Spin-Orbit-Coupled Mott Insulator}

\author{Alexis Ghermaoui}
\affiliation{Laboratoire Kastler Brossel, Collège de France, CNRS, ENS-Université PSL,
Sorbonne Université, 11 place Marcelin Berthelot, F-75231 Paris, France}

\author{Manel Bosch Aguilera}
\affiliation{Laboratoire Kastler Brossel, Collège de France, CNRS, ENS-Université PSL,
Sorbonne Université, 11 place Marcelin Berthelot, F-75231 Paris, France}

\author{Rapha\"el Bouganne}
\affiliation{Laboratoire Kastler Brossel, Collège de France, CNRS, ENS-Université PSL,
Sorbonne Université, 11 place Marcelin Berthelot, F-75231 Paris, France}

\author{R\'emy Vatr\'e}
\affiliation{Laboratoire Kastler Brossel, Collège de France, CNRS, ENS-Université PSL,
Sorbonne Université, 11 place Marcelin Berthelot, F-75231 Paris, France}

\author{Isabella Fritsche}
\affiliation{Institut f{\"u}r Quantenoptik und Quanteninformation (IQOQI),
{\''O}sterreichische Akademie der Wissenschaften, 6020 Innsbruck, Austria}
\affiliation{Institut f{\"u}r Experimentalphysik, Universit{\"a}t Innsbruck, 6020 Innsbruck, Austria}

\author{J\'er\^ome Beugnon}
\affiliation{Laboratoire Kastler Brossel, Collège de France, CNRS, ENS-Université PSL,
Sorbonne Université, 11 place Marcelin Berthelot, F-75231 Paris, France}

\author{Fabrice Gerbier}
\email[Corresponding author:]{fabrice.gerbier@lkb.ens.fr}
\affiliation{Laboratoire Kastler Brossel, Collège de France, CNRS, ENS-Université PSL,
Sorbonne Université, 11 place Marcelin Berthelot, F-75231 Paris, France}


\date{\today}

\begin{abstract}
We use Ramsey interferometry to study spin dynamics in the strongly interacting regime of spin-orbit-coupled quantum gases in one-dimensional optical lattices. We observe an intrinsic many-body dephasing mechanism  immune to spin-echo in two-component Mott insulators. We ascribe the dephasing to the motion of hole-like defects in an otherwise inert Mott insulator, the spinless nature of the holes explaining the ineffectiveness of spin echo to restore it.  We show that a model of spin-orbit-coupled hardcore bosons can explain quantitatively our experimental observations.
\end{abstract}

\maketitle
%
%

%
%
%
%

Spin-orbit interactions coupling together the internal and external (motional) degrees of freedom of a quantum particle play a key role in modern condensed matter physics, from spintronics to topological insulators\,\cite{hasan2010a,qi2011a}. In isolated atoms or solids, spin-orbit interactions come from relativistic correction to the motion of electrons. In quantum gases of neutral atoms\,\cite{bloch2008a}, they can be engineered using external radiation coupling two internal states $g,e$ identified with the $\uparrow,\downarrow$ component of a pseudo-spin-1/2 particle\,\cite{lin2011a, goldman2014a,zhai2015a,li2015a,cooper2019a}. The recoil momentum associated with the absorption of a photon when the atom is promoted from $g$ to $e$ provides the desired coupling between degrees of freedom. 

This general principle has been explored using two-photon transition coupling two hyperfine states in the ground manifold of alkali atoms\,\cite{lin2011a,wang2012a,cheuk2012a, qu2013a,huang2016a,li2017a,kroeze2019a} or one-photon transition between a ground state level and a metastable excited state (a so-called ``clock state'' with negligible radiative lifetime)\,\cite{livi2016a,wall2016a,song2016a,kolkowitz2017a,bromley2018a,aeppli2022a}. The latter situation is also of great interest for time-frequency metrology as building block for the most accurate atomic clocks to date that operate in the optical domain\, \cite{ludlow2015a,kennedy2020a,bothwell2022a,young2020a,kim2023a}.
Far from the traditional ``single atom'' picture prevalent in previous generations of atomic clocks, experiments in this context have explored two-\,\cite{campbell2009a,zhang2014a,goban2018a,hutson2023a}
and many-body effects\,
\cite{martin2013a,zhang2014a,rey2014a,campbell2017a,marti2018a,hutson2019a}. In combination with optical lattices\,\cite{bloch2008a}, quantum gases of atoms with such an optical clock transition  offer an exceptionally rich experimental platform where quantum coherence, interactions and spin-orbit coupling are all paramount and where many physical situations previously studied for two-component gases without spin-orbit coupling\, \cite{duan2003a,altman2003a,garciaripoll2004b, vengalattore2008a,hild2014a,nichols2019a,venegasgomez2020a,jepsen2020a,
jepsen2021a} are expected to change qualitatively\,\cite{struck2013a,hamner2015a,zhang2015a,yamamoto2017a}. Furthermore, several proposals highlight the unique advantages of this platform to explore novel quantum many-body phases or to
generate and harness entanglement\,\cite{inaba2014a,he2019a,mamaev2019a,gonzalezcuadra2023a}.

\begin{figure}[t!!!!!]
\centering
\includegraphics[width=\columnwidth]{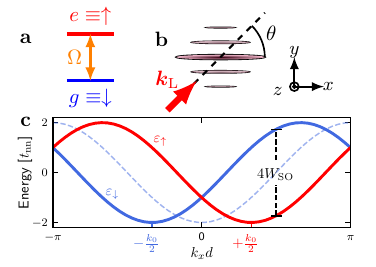}
\caption{\textbf{a}:  Sketch of the internal state scheme and \textbf{b}: of the experimental geometry. The projection of the wavevector $\bmkL$ of the coupling laser on the $x$ axis of the one-dimensional gases  determines the recoil momentum $k_0=\kL \cos \theta$ imparted to each atom promoted to the state $\uparrow$. \textbf{c}: Single-particle energy dispersion relations in the spiral frame, $\epsilon_{\uparrow/\downarrow}(k_x) = -2 t_{\mathrm{n.n.}} \cos\big[(k_x \mp k_0 /2) d\big)]$. The dashed line shows the dispersion relation without spin-orbit coupling, $\epsilon(k_x) = -2 t_{\mathrm{n.n.}} \cos\big(k_x d)$.
}
\label{fig1}
\end{figure}

In this work, we study strongly-interacting quantum gases of spin-orbit-coupled  two-component bosons in one-dimensional optical lattices.  Using the clock transition  of $^{174}$Yb atoms, we perform Ramsey interferometry\,\cite{abanin2013a,knap2013a} to characterize the spin dephasing due to many-body dynamics in the Mott insulator regime. We find that this dephasing can be ascribed to spin-orbit coupling and
is immune to spin echo, in stark contrast with the previously studied single-particle case\,\cite{kolkowitz2017a}. Using a model of hardcore bosons, we show that the motion of hole-like defects in an otherwise inert Mott insulator explains the measured loss of contrast and the ineffectiveness of spin echo to restore it. Experimentally, such defects could originate from finite temperature of the sample, deviations from adiabaticity in the preparation stage or inelastic losses when the metastable excited state is populated.

Our experiments are performed with arrays of independent one-dimensional lattice gases. We load a nearly pure Bose-Einstein condensate of $^{174}$Yb atoms in a cubic but strongly anisotropic optical lattice. The lattice depths in the $y/z$ directions are set at their maximum value  $V_{0y/z} \simeq 26\,\ER$ to suppress tunneling, with  $\ER\simeq h \times 1.5\,$kHz the recoil energy at the lattice wavelength $\lambda_{\mathrm{lat}} =2d 759.9\,$nm ($d$ is the lattice spacing).  The lattice depth  $V_{0x}  $ in the $x$ direction varies in the range $V_x = 6-26\,\ER$, which corresponds to the Mott insulator regime in one dimension. We work with an atom number around $\sim 8 \cdot 10^3$, such that the shell structure in a harmonic trap\,\cite{bloch2008a} consists of a single unit-filled Mott plateau.

The two internal states of interest are the electronic ground state $^1$S$_0\equiv \downarrow $ and the clock state $^3$P$_0\equiv \uparrow $. We prepare an initial state with all atoms in the $\downarrow$ state. A near-resonant laser of wavelength $2\pi/\kL \approx 578\,$nm  provides the $\downarrow \leftrightarrow \uparrow$ coupling. A convenient choice to describe the dynamics is the so-called ``spiral frame'', where the recoil momentum is absorbed in the definition of the single-particle states, $\vert \sigma \rangle_m \to \ee^{-\ii \sigma k_0 d m/2)}\vert \sigma \rangle_m $ (see the Supplemental Material\,\cite{SM} for details). Here and below we use the notation $\sigma$ either for the state labels or for their ``signature'' $\sigma=\pm 1$ for $\uparrow/\downarrow$,  and denote by $\hat{a}_{m,\sigma}$, $\hat{s}_{m,\alpha}$ and $\hat{S}_\alpha=\sum_m \hat{s}_{m,\alpha}$ ($\alpha=x,y,z$) the  annihilation operator of a spin-$\sigma$ boson at site $m$ and the local and total spin operators, respectively.

The Hamiltonian in the spiral frame takes the form $\hat{H} = \hat{H}_{0} -\hbar \deltaL \hat{S}_z+ f(t)\hat{V}_{\mathrm{L}} $, with a motional term $\hat{H}_{0} = \hat{K}   + \hat{H}_{\mathrm{int}}$ regrouping the single-particle and interaction Hamiltonians, a Zeeman-like term that depends on the detuning  $\deltaL =\omegaL-\omega_{0}$ of the laser frequency $\omegaL$ from the atomic resonance frequency $\omega_0$, and an electric dipole  term $\hat{V}_{\mathrm{L}} $ coupling $\uparrow \leftrightarrow \downarrow$. The pulse function $f(t)$ is equal to one (zero) when the coupling laser is turned on (off). The dispersion relations for the one-dimensional single-particle tight-binding  Hamiltonian  
\begin{align}
\label{eq:Kspiral}
\hat{K} =& -t_{\mathrm{nn}}\sum_{m, \sigma } \ee^{-\ii \sigma \frac{k_0 d}{2}} \hat{a}_{m+1,\sigma}^\dagger \hat{a}_{m,\sigma} + \mathrm{h.c.},
\end{align} 
with $t_{\mathrm{nn}}$ the nearest-neighbor tunneling energy, are shown in Figure\,\ref{fig1}c. Spin-orbit coupling induces a $k_x$-dependent Zeeman shift $\varepsilon_\uparrow-\varepsilon_\downarrow = -4 W_{\mathrm{SO}} \sin(k_x d) $ with the ``spin-orbit coupling strength'' 
\begin{align}
\label{eq:WSO}
W_{\mathrm{SO}} &  = t_{\mathrm{nn}} \sin\left(\frac{k_0 d}{2}\right).
\end{align}
 Our experimental parameters correspond to $\kL d \simeq 1.313 \,\pi$.  The electric dipole Hamiltonian in the spiral frame reads $\hat{V}_{\mathrm{L}}(\phi)  =  -\hbar \Omega   \bm{\hat{S}} \cdot \bm{v}(\phi)$ with a Rabi frequency $\Omega$ and a rotation axis $\bm{v}(\phi) = -\sin \phi \bm{e}_x + \cos \phi\bm{e}_y$, with $\phi$ the laser phase \,\cite{SM}.

\begin{figure}[ht!!!]
\centering
\includegraphics[width=\columnwidth]{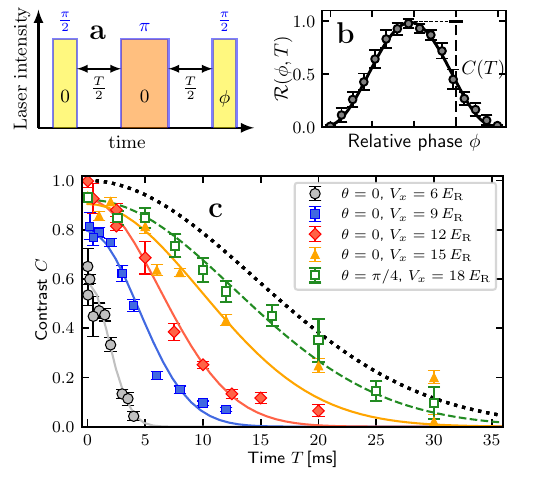}  
\caption{\textbf{a}: Pulse timing of a two-axis Ramsey sequence with spin echo. \textbf{b}: Typical Ramsey fringe for a Ramsey time $T=200\,\mu$s and a lattice depth $V_x=26\,\ER$. The Rabi frequency characterizing the driving strength is $\Omega/(2\pi) = 1389\,$Hz, corresponding to a $\pi/2$ pulse time $T_{\mathrm{p}} \simeq 250\,\mu$s. \textbf{c}: Decay of the Ramsey fringe contrast with the Ramsey  time $T$ for several lattice depths $V_x =6-18\,\ER$ and incidence angles of the coupling laser $\theta=0, \pi/4$ (see legend). The solid lines show the fits to the experimental data using Eq.\,(\ref{eq:R}). The black dotted line is the contribution $\ee^{-F_{\mathrm{SE}}}$ from the laser frequency noise. Error bars are the statistical standard deviations of each measurement, with typically $5$ measurements per data point. }
\label{fig2}
\end{figure}

We probe the spin dynamics of the  one-dimensional lattice gases realized in the experiment with Ramsey interferometry. A time-domain Ramsey interferometer uses two successive $ \pi/2$ pulses of duration $\pi/(2\Omega)$ separated by a laser-free evolution time of duration $T $. The first ``preparation pulse'' with $\phi=0$ maps the state $\vert \downarrow \rangle_m$ to the balanced superposition state $\vert +\bm{x}\rangle_m = \left( \vert \downarrow \rangle_m  + \vert \uparrow \rangle_m \right)/\sqrt{2}$, the  $+ 1/2$ eigenstate of $\hat{s}_{m,x}$. A  second  ``readout pulse'' with $\phi \neq 0$ closes the interferometer, so that the normalized  population in the state $\downarrow$ (``Ramsey signal''  hereafter),
\begin{align}
\label{eq:R}
\mathcal{R}(\phi,T)  = \left\langle \frac{1}{2}-\frac{ \hat{S}_{z} }{N}\right\rangle =  C(T) \cos(\zeta+\phi_0-\phi),
\end{align} 
 varies sinusoidally with the phase difference $\zeta+\phi_0-\phi$ accumulated between the two ``paths'' $\downarrow/\uparrow$. Here the dynamical phase due to the Zeeman-like Hamiltonian $\zeta=\omega_0 T$ is equivalent to a ``Larmor phase'' in the pseudo-spin-1/2 picture, $\phi_0(T)$ is an eventual additional contribution due to atomic motion or interactions, and $\phi$ is the relative phase of the two waves driving the first and second pulses. 

While Ramsey interferometry is often (\textit{e.g.} in atomic clocks) used to perform precision measurements of the Larmor phase, the signal of interest in most quantum gases experiments (and in this work in particular) is the evolution of the contrast $C(T)$ due to many-body dynamics. In this context, it is essential to be able to distinguish the intrinsic system dynamics from extrinsic contributions that also lead to a contrast decay, \textit{e.g.} due to spatial inhomogeneities or temporal fluctuations of the Larmor phase. In order to mitigate the extrinsic effects, we use a spin echo sequence where a $\pi$ pulse is inserted in the middle of the sequence (Fig.\,\ref{fig2}\textbf{a}). 

\begin{figure}[h!!!]
\centering
\includegraphics[width=\columnwidth]{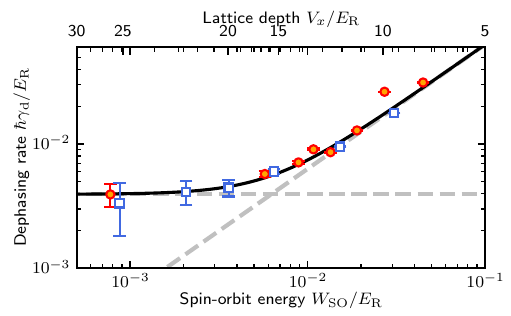}
\caption{ Dephasing rate $\gamma_{\mathrm{d}}$ versus spin-orbit energy $W_{\mathrm{SO}}$ [lattice depth $V_x$ in the top axis] and two different incidence angles, $\theta=0$ ($ \varphi_{\mathrm{rec}} \approx 1.37 \pi$ -- filled circles) and $\theta=\pi/4$ ($\varphi_{\mathrm{rec}} \approx 0.97\, \pi$ -- open squares). The solid line shows the fit by Eq.\,(\ref{eq:gammad}), while the dashed lines show the asymptotes for small or large $W_{\mathrm{SO}}$. Error bars are the $1/e^2$ confidence intervals of the fits to the contrast curves estimated using standard error propagation.  }
\label{fig3}
\end{figure}

We measure the Ramsey signal with spin echo as a function of the Ramsey time $T$ and for several lattice depths $V_x$ and two different angles of the coupling laser $\theta=0,\pi/4$. For given $V_x,T$, we record Ramsey fringes as a function of $\phi$ and extract their contrast $C(T)$ by fitting the experimental curves to Eq.\,(\ref{eq:R}) [see Fig.\,\ref{fig2}\textbf{b}]. A representative subset of the experimental data is shown  in Fig.\,\ref{fig2}\textbf{c}. In turn, we fit the measured contrast with a Gaussian function,
 \begin{align}
C(T) &  =C_0 \exp\left[ - \big(\gamma_{\mathrm{d}} T\big)^2 \right],
\end{align}
with the initial contrast $C_0$ (which is close to unity) and the dephasing rate $\gamma_{\mathrm{d}}$ as free parameters (the choice of a Gaussian function  will be discussed below). The fitted dephasing rates are shown in Fig.\,\ref{fig3} versus the spin-orbit coupling strength $W_{\mathrm{SO}}$. We find that the dephasing rate is well described by the law
 \begin{align}
 \label{eq:gammad}
\gamma_{\mathrm{d}} &  = \sqrt{ \gamma_{\mathrm{L}}^2 + \left(  \frac{\alpha W_{\mathrm{SO}}}{\hbar} \right)^2},
\end{align}
with a baseline value $\gamma_{\mathrm{L}}=49(1)$\,s$^{-1}$ and $\alpha \approx 0.63(1)$ a numerical coefficient. The baseline represents the residual effect of the frequency fluctuations of the coupling laser that are only partially compensated by the echo pulse. 

To understand the experimental results, we first show that under experimentally sensible assumptions, the Ramsey signal can be expressed in a simple form where the Ramsey phase $\phi$, the laser frequency fluctuations, and the many-body dynamics during the Ramsey time $T$ are clearly separated. To take the frequency noise of the coupling laser into account, we treat the detuning $\deltaL(t)$ as a Gaussian stochastic process centered at zero. We assume (\textit{i}) that  $\hat{H}_{\mathrm{int}}$ is invariant under spin rotations, $[\hat{H}_{\mathrm{int}}, \hat{S}_{x/y/z}]=0$. Since $ [\hat{K}, \hat{S}_{x/y/z} ]=0$, the total spin operators are constants of motion between the pulses. To make the theoretical analysis tractable, we also assume \textit{(ii)} that the coupling $\hat{V}_L$ is much stronger than $\hat{H}_0$. Formally, we take the idealized limit $\Omega \to \infty, T_{\mathrm{p}} \to 0$ with $\Omega T_{\mathrm{p}}=\pi/2$, with $T_{\mathrm{p}}$ the pulse duration. The evolution operator corresponding to the preparation pulse (respectively, read-out pulse) is then well approximated by the spin rotation operator $\hat{R}_1 = \exp(-\ii \hat{V}_{\mathrm{L}}(0) T_{\mathrm{p}}/\hbar)$ [resp., $\hat{R}_2 = \exp(-\ii \hat{V}_{\mathrm{L}}(\phi) T_{\mathrm{p}}/\hbar)$]. The evolution operator during the much longer Ramsey evolution time contains the genuine many-body dynamics. 
Using the two assumptions \textit{(i)},\textit{(ii)} and standard angular momentum algebra, we derive a compact formula for the Ramsey signal with spin echo\,\cite{SM}
\begin{align} 
\label{eq:R_SE}
\mathcal{R}(T,\phi) & = \frac{1}{2}+\ee^{-F_{\mathrm{SE}}(T)}  \langle \bm{\hat{S}} \cdot \bm{u}(\phi)\rangle_{T}.
\end{align}
The read-out pulse measures the spin projection along the axis $\bm{u} (\phi)= \cos \phi \bm{e}_x + \sin \phi\bm{e}_y$, with a corrective term $\ee^{-F_{\mathrm{SE}}} $ accounting for the residual effect of the laser frequency fluctuations. The spin expectation values
$  \langle \bm{\hat{S}}_\alpha \rangle_{T}  = (1/N)\sum_m  \langle \Psi(T) \vert  \hat{S}_{m,\alpha} \vert \Psi(T) \rangle$ are solely determined by the time-evolved many-body state
 $\vert \Psi(T) \rangle = \hat{U}_{\mathrm{SE}}(T)\hat{R}_1 \vert \Psi(0) \rangle$, with $\hat{U}_0(T)  = \exp(-\ii \hat{H}_0 T/\hbar)$ and $\hat{U}_{\mathrm{SE}}(T) = \hat{U}_0(T/2)  \hat{R}_1^2 \hat{U}_0(T/2)$ the evolution operators without or with spin echo, respectively. Conversely, the noise factor $\ee^{-F_{\mathrm{SE}}} $ is determined by the spectrum of frequency fluctuations and the pulse length and shape, \textit{independently of the many-body evolution}. Eq.\,(\ref{eq:R_SE}) holds for bosons and fermions, irrespective of the exact form of $\hat{H}_{\mathrm{int}}$ provided it is invariant under spin rotations.

\begin{figure}[h!!!]
\centering
\includegraphics[width=\columnwidth]{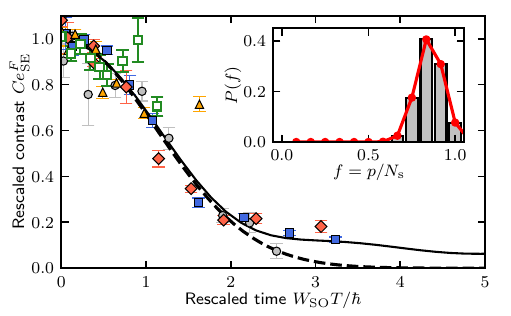}
\caption{Rescaled Ramsey contrast for several lattice depths $V_x$ and incidence angles  $\theta$ of the coupling laser: $\theta=0$, $V_x =6\,\ER$ (circles), $\theta=0$, $V_x =9\,\ER$ (filled squares), $\theta=0$, $V_x =12\,\ER$ (diamonds), $\theta=0$, $V_x =15\,\ER$ (triangles) and $\theta=\pi/4$, $V_x =18\,\ER$ (open squares). The horizontal axis is rescaled by the relevant frequency scale $W_{\mathrm{SO}}/\hbar$ in presence of spin-orbit coupling and the contrast curves by $C_0 \ee^{-F_{\mathrm{SE}}}$, with $C_0$ the fitted initial valued and $F_{\mathrm{SE}}$ the calibrated noise factor (see text).  Error bars are the statistical standard deviations of each measurement. The solid line shows a calculation for two-component hardcore bosons in a lattice with $N_s=12$ sites, where an initial state taken to be a statistical mixture of Fock states with weights $P(f)$ shown in the inset ($p$ the number of bosons). }
\label{fig4}
\end{figure}

Using Eq.\,(\ref{eq:R_SE}), we normalize away the effect of laser noise and extract the genuine many-body dynamics from the data. We  use a procedure similar in spirit to the ``optical noise analyzer'' demonstrated in \cite{bishof2013a} to calibrate the laser noise factor $\ee^{-F_{\mathrm{SE}}} $. For this calibration, we work with an atomic gas held in a very deep and almost isotropic lattice of depth $ \simeq26\,\ER$ where tunneling is negligible. We find\,\cite{SM} that the noise factor is well approximated by a Gaussian function,  $\exp[-F_{\mathrm{SE}}(T)] \simeq \exp[-(\gamma_ {\mathrm{\Delta\wL}} T)^2]$, with $\gamma_ {\mathrm{\Delta\wL}} \approx 49(3)\,$s$^{-1}$,) a value consistent with the baseline dephasing rate $\gamma_{\mathrm{L}}$ determined earlier. For experimental data with $V_x \in [6-18]\,\ER$, we rescale each measured contrast by $C_0\ee^{-F_{\mathrm{SE}}} $ to remove the effect of laser noise and time by $\hbar/W_{\mathrm{SO}}$. After rescaling, the data collapse on a single ``universal'' curve   shown in Figure\,\ref{fig4} governed by the spin-orbit timescale $\hbar/W_{\mathrm{SO}}$. The observation of rescaling provides further evidence for the claim that the observed dephasing originates from motional effects in presence of spin-orbit coupling.

We now discuss the role of atomic motion through the lattice \textit{via}  quantum tunneling and its impact on the Ramsey signal. For a single particle initially prepared in a particular Wannier state (and, by extension, for an ensemble of initially well-localized but non-interacting particles), the Ramsey signal can be calculated analytically\,\cite{kolkowitz2017a},
\begin{align}
C(T) &= \begin{cases}
J_0 \left( \frac{4 W_{\mathrm{SO}} T}{\hbar} \right)& (\mathrm{no\,\,\,spin\,\,\,echo)},\\
1& (\mathrm{spin\,\,\,echo)},
\end{cases}
\end{align}
with $J_0$ a Bessel function of the first kind. Without spin echo, dephasing arises from a momentum-dependent Zeeman shift, the analog of Doppler broadening for a quantum particle moving on a tight-binding lattice. This broadening is entirely undone by the spin echo pulse and the spin-echo Ramsey signal \textit{for non-interacting particles} is thus unaffected by motional effects\,\cite{kolkowitz2017a}. Thus, atomic motion \textit{alone} cannot explain our observations. 

\begin{figure}[ht!!!]
\centering
\includegraphics[width=\columnwidth]{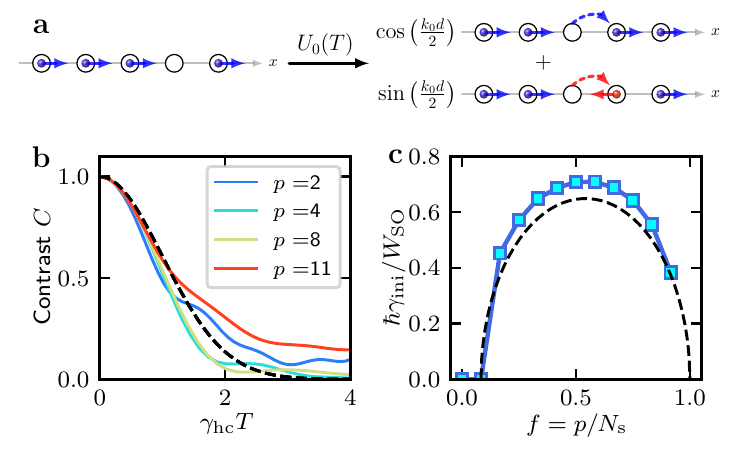}
\caption{ Numerical simulations for $p$ hardcore bosons in a one-dimensional lattice with $N_s=12$ sites. \textbf{a.} Sketch of the evolution through quantum tunneling  of an initial state with a single hole  tunneling. \textbf{b.} Calculated contrast with spin echo against rescaled time for several filling factors $f=p/N_s$, where the characteristic rate $\gamma_{\mathrm{hc}}$ is defined in Eq.\,(\ref{eq:gammahc}). \textbf{c.} Dephasing  $\gamma_{\mathrm{ini}}$ versus lattice filling fraction $\bar{n}=p/N_s$ (squares). The dashed lines in \textbf{b,c} show $\exp(-[\gamma_{\mathrm{hc}} T]^2)$ and $\gamma_{\mathrm{hc}}$, respectively. The simulations are performed with $k_0 d=4\pi/3$ and periodic boundary conditions. }
\label{fig5}
\end{figure}

In the opposite limit of a unit-filled Mott insulator, tunneling of particles is energetically blocked by a strong on-site repulsion $U$. Virtual transitions to doubly-occupied states lead to an effective spin-spin (``superexchange'') interaction with a characteristic exchange energy $J_{\mathrm{ex}} \sim \tnn^2/U \ll \tnn$. On the time scale $\hbar/W_{\mathrm{SO}}$, superexchange has a negligible influence and cannot explain the observed dephasing. For small deviations from unit filling, the initial state can be pictured as a unit-filled MI with mobile hole-like defects. To understand how the defects impact spin coherence, it is useful to rewrite the tunneling Hamiltonian in the $\pm  \bm{x}$ basis,
\begin{align}
\nonumber
\hat{K} =&\sum_{m, \sigma = \pm \bm{x},d=\pm 1 }  -\overline{W} \hat{a}_{m+d,\sigma}^\dagger \hat{a}_{m,\sigma}\\
\label{eq:Kspiral2}
& \hspace{2cm}  + \ii \,d  W_{\mathrm{SO}} \hat{a}_{m+d,-\sigma}^\dagger \hat{a}_{m,\sigma},
\end{align}
with $\overline{W}=\tnn \cos(k_0d/2)$. Thus, a tunneling ``event'' leads to a superposition state illustrated in Fig.\,\ref{fig5}a, where a hole moving to a neighboring site leaves behind a ``spin-flip'' defect with a probability amplitude $\propto W_{\mathrm{SO}}$. 

To describe quantitatively this hole-induced dephasing mechanism, the many-body problem can be formulated as a bosonic $\tnn-J_{\mathrm{ex}}$ model, which captures hole tunneling and superexchange interactions at the same time. Because superexchange has a negligible impact, we simplify the theory to its minimal version and consider  a model of two-component hardcore bosons defined by the kinematic relations
$a_{m,\sigma}a_{m,\sigma'}=0$, which forbid double occupancies, and by the Hamiltonian $\hat{H}_0 = \hat{K}$. 
From Eq.\,(\ref{eq:R_SE}) without frequency fluctuations, the Ramsey contrast is given by $C = 2\vert \langle \hat{S}_x \rangle_{\mathrm{f}}\vert$. We rely on exact diagonalization for small systems with $N_s=12$ lattice sites to compute the spin expectation value. We assume an initial Fock state with $p \in [1,11]$ particles and call $f = p/N_s$ the filling factor. A few illustrative examples of the calculated contrast $C(T)$ are shown in Fig.\,\ref{fig5}b. We find that the initial decay is reasonably well approximated by a Gaussian function $C(T,f) \simeq \exp\big[-(\gamma_{\mathrm{ini}}T)^2\big]$. The fitted Gaussian dephasing rate $\gamma_{\mathrm{ini}}$ is well reproduced by the heuristic formula [Fig.\,\ref{fig5}c]
\begin{align}\label{eq:gammahc}
\frac{\hbar\gamma_{\mathrm{hc}}}{W_{\mathrm SO}} =   \sqrt{2 (f-N_s^{-1})(1-f)} .
\end{align}
The form in Eq.\,(\ref{eq:gammahc}) ensures that the dephasing rate vanishes for a single particle ($f=1/N_s$) and reproduces a perturbative analytical result for a single hole ($f=1-1/N_s $)\,\cite{SM}. 

To compare with the experimental data, we model the real initial state as a statistical mixture of Fock states with weights $P(f)$, and the weight distribution $P(f)=N \ee^{-(f-\bar{f})^2/(2\Delta f^2)}$ as a truncated Gaussian distribution of width $\Delta f$ centered at $\bar{f}$. We compare the numerically calculated Ramsey decay curves averaged over $P(f)$ to the rescaled data in Fig.\,\ref{fig4}.  With the best fit parameters $\bar{f} \simeq 0.77$ and $\delta f \simeq 0.08$  (the corresponding $P(f)$ is shown in the inset), we find that the hardcore bosons model, shown as solid line, reproduces well the initial Gaussian decay as well as the slower tail of the curve. The dashed line in Fig.\,\ref{fig4}a is the Gaussian function $\exp\big[-(\alpha W_{\mathrm{SO}}T/\hbar)^2\big]$ with the  experimentally determined value $\alpha \approx 0.63(1)$. The deviations from a Gaussian at longer times express the fact that the motion of a hole at long times is perturbed by the string of defects left behind by its own motion or by the motion of other holes and becomes more complex than the initial ballistic flight. 

In conclusion, we have reported on experiments using Ramsey interferometry to probe the dynamics of spin-orbit-coupled 1D lattice gases in the strongly interacting regime. We have identified an intrinsic dephasing mechanism impervious to spin echo (and by extension to other dynamical decoupling techniques\,\cite{lidar2014a}) that we ascribe to the motion of spinless hole-like defects on top of a background of spinful but inert atoms. We believe that this mechanism limits the coherence times of optical lattice atomic clocks operating in the Mott insulator regime, where a non-zero defect density  inevitably occurs to some level. Quite generally,  the topic of hole motion against various magnetic backgrounds is highly relevant to understand strongly-correlated materials\,\cite{carlstroem2016a,grusdt2018a,ji2021a,bohrdt2021a}. Extending our work to study single 1D  wires or 2D planes in combination with high-resolution imaging would enable a quantitiative comparison between the various theories put forward in such a context and experiments.  

\begin{acknowledgements}
We thank the members of the BEC group at LKB for various stimulating discussions.
\end{acknowledgements}

\bibliography{BibRamsey}
\bibliographystyle{apsrev}

\end{document}